\documentclass{cambridge6A}
  \usepackage{graphicx}
  \usepackage{amssymb,amsmath,amsfonts}

\begin{document}

\def \lsim{\mathrel{\vcenter
     {\hbox{$<$}\nointerlineskip\hbox{$\sim$}}}}
\def \gsim{\mathrel{\vcenter
     {\hbox{$>$}\nointerlineskip\hbox{$\sim$}}}}

\newcommand{\beq}{\begin{equation}}
\newcommand{\eeq}{\end{equation}}
\newcommand{\be}{\begin{equation}}
\newcommand{\ee}{\end{equation}}
\newcommand{\beqa}{\begin{eqnarray}}
\newcommand{\eeqa}{\end{eqnarray}}
\newcommand{\beqar}{\begin{eqnarray*}}
\newcommand{\eeqar}{\end{eqnarray*}}
\newcommand{\eg}{{\it e.g.,}\ }
\newcommand{\ie}{{\it i.e.,}\ }
\newcommand{\vep}{\varepsilon}
\newcommand{\labell}[1]{\label{#1}} 
\newcommand{\labels}[1]{\label{#1}} 
\newcommand{\reef}[1]{(\ref{#1})}
\newcommand{\bea}{\begin{eqnarray}}
\newcommand{\eea}{\end{eqnarray}}
\newcommand{\oln}{\overline\nabla}
\newcommand{\Ld}{\pounds}

\renewcommand*\thesection{\arabic{section}}
\renewcommand*\thesubsection{\arabic{section}.\arabic{subsection}}
\setcounter{equation}{0}
\numberwithin{equation}{section}
\setcounter{figure}{0}
\renewcommand{\thefigure}{\arabic{figure}.}

\chapterauthor{Roberto Emparan \\ ~ \\
\textit{Instituci\'o Catalana de Recerca i Estudis Avan\c cats (ICREA)\\ and 
\\ Departament de F{\'\i}sica Fonamental and Institut de
Ci\`encies del Cosmos, Universitat de Barcelona, Mart\'{\i} i Franqu\`es
1, Barcelona, Spain}}

\chapter*{Blackfolds}

\contributor{Roberto Emparan
\affiliation{Instituci\'o Catalana de Recerca i Estudis Avan\c cats
(ICREA), Passeig Llu\'{\i}s Companys 23, E-08010 Barcelona, Spain} 
\affiliation{Departament de F{\'\i}sica Fonamental and Institut de
Ci\`encies del Cosmos, Universitat de
Barcelona, Mart\'{\i} i Franqu\`es 1, E-08028 Barcelona, Spain}
}

\begin{abstract}\small
This is an introduction to the blackfold effective worldvolume theory
for the dynamics of black branes, as well as its use as an approximate
method for the construction of new black hole solutions. We also explain
how the theory is useful for the analysis of dynamical, non-stationary
situations, in particular of the Gregory-Laflamme instability of black
branes.
\end{abstract}

\copyrightline{Chapter of the book \textit{Black Holes in Higher Dimensions} to
be published by Cambridge University Press (editor: G. Horowitz)}

\section{Introduction}

The existence of black $p$-branes in higher-dimensional General
Relativity hints at the possibility of large classes of black holes
without any four-dimensional counterpart. Black rings provide a nice
explicit example: in \cite{chap:blackrings} they were introduced as
the result of bending a black string into the shape of a circle and
spinning it up to balance forces. One can naturally expect that this
heuristic construction extends to other black branes. If the worldvolume
of a black $p$-brane could be similarly bent into the shape of a compact
hypersurface, for instance a torus $\mathbb{T}^p$ or a sphere $S^p$, we
would obtain many new geometries and topologies of black hole horizons. 

Unfortunately, the techniques that allow to construct \textit{exact}
black hole solutions in four and five dimensions have not been
successfully extended to more dimensions. Still, one may want to hold on
to the intuition that a long circular black string, or more generally a
smoothly bent black brane, could be obtained as a perturbation of a
straight one. 

The experience with brane-like objects in other areas of physics
suggests that such approximate methods may be efficiently applied to
this problem. Consider, for instance, the Abelian Higgs theory and its
familiar string-like vortex solutions. These are first obtained in the
form of static, straight strings, but one expects that they can also
bend and vibrate. It has been long recognized that, if the wavelength of
these deformations is much longer than the thickness of the vortex, the
dynamics of the full non-linear theory is well approximated by the
simpler Nambu-Goto worldsheet action. One can then use it for studying
loops of strings of diverse shape. Another example is provided by
D-branes in string theory, which are defined as surfaces where open
strings can attach their endpoints. Although the bending and vibrations
of D-branes are generally intractable in an exact manner in string
theory, they are again very efficiently captured by the
Dirac-Born-Infeld worldvolume field theory, which is applicable as long
as the scale of the deformations is sufficiently large that locally the
brane can still be well approximated as a flat D-brane. In all these
cases, the full dynamics of the brane is replaced by an effective
worldvolume theory for a set of ``collective coordinates''.

So, similarly, we seek an effective theory for describing black branes
whose worldvolume is not exactly flat, or not in stationary equilibrium,
but where the deviations from the flat stationary black brane occur
on scales much longer than the brane thickness. Black branes whose
worldvolume is bent into the shape of a submanifold of a
background spacetime have been named \textit{blackfolds}. 
 
This chapter introduces the blackfold effective worldvolume theory for
the dynamics of black branes, as well as its use as an approximate
method for the construction of new black hole solutions
\cite{Emparan:2009cs,Emparan:2009at,Emparan:2009vd}. Of special interest
is a class of helical black rings that provide the first example of
black holes in all $D\geq 5$ with the minimum of symmetry required by
rigidity theorems. We also explain how the theory is useful for the
analysis of dynamical, non-stationary situations, in particular of the
Gregory-Laflamme instability of black branes reviewed in
\cite{chap:GLinstab}. The blackfold techniques connect it very directly
to the instability of an effective ``black brane fluid'', in a manner
that shares features with the fluid/gravity correspondence of
\cite{chap:fluidgrav}.

A word about notation in this chapter. When considering a $p$-brane with
worldvolume
$\mathcal W_{p+1}$ embedded in a $D$-dimensional background spacetime,
we denote
\beq\label{defn}
n=D-p-3\,.
\eeq 
Spacetime indices $\mu,\nu$ run in $0,\dots,D$ and the covariant
derivative compatible with the background metric $g_{\mu\nu}(x)$ is
$\nabla_\mu$. Worldvolume indices $a,b$ run in $0,\dots,p$, and the
covariant derivative compatible with the metric $\gamma_{ab}(\sigma)$ induced on
$\mathcal W_{p+1}$ is $D_a$.

\section{Effective theory for black hole motion}

Above we have motivated the blackfold effective approach by drawing
analogies between black branes and the extended brane-like solutions of
other non-linear theories. However, the general-relativistic aspects of
the effective theory of black $p$-brane dynamics are better introduced
by considering first the simpler case of $p=0$: the effective dynamics
of a black hole that moves in a background whose curvature radii $\sim
R$ are large compared to the black hole horizon radius $r_0$,
\beq\label{sepscales}
r_0\ll R\,.
\eeq

This separation of scales implies the existence of two distinct regions
in the geometry. First, there is
a region around the black
hole where the geometry is well approximated by the Schwarzschild(-Tangherlini)
solution. If in this  ``near zone'' we choose a coordinate $r$ centered at the black
hole, the Schwarzschild geometry is a good approximation as long as
$r\ll R$, i.e.,
\beq
ds^2_{(near)}=ds^2(\mathrm{Schwarzschild})+O(r/R)\,.
\eeq
The corrections to the Schwarzschild metric are the (tidal) distortions
that the background
curvature creates on the black hole. 

The second region is far enough from the black hole that its effect
on the background geometry is very mild and
can be treated as a small
perturbation. This is the ``far zone'' where $r\gg r_0$,
and in which we can expand 
\beq
ds^2_{(far)}=ds^2(\mathrm{background})+O(r_0/r)\,.
\eeq
Since we are too far from the black hole to resolve its size,
effectively it is a pointlike
source whose gravitational effect on the background can be computed
perturbatively. To
this source we can assign an effective trajectory
$X^\mu(\tau)$,
with proper time $\tau$ and with velocity $\dot X^\mu=\partial_\tau X^\mu$
such that $g_{\mu\nu}\dot X^\mu \dot X^\nu=-1$. We also
ascribe to it an effective stress-energy tensor that encodes how the black
hole affects the gravitational field in the far zone.

Let us determine the general form of this stress-energy tensor. We can
naturally assume that the acceleration and other higher derivatives of the
particle's velocity are small, since they must be caused by the deviations away
from flatness of the background in the region where the black hole
moves. Since these variations occur on scales $\sim R$, these
higher-derivative terms must be suppressed by powers of
$r_0/R$. To leading
order in this expansion, the stress-energy tensor of the effective
source is fixed by symmetry and worldline
reparametrization invariance to have the form
\beq\label{efftmunu}
T^{\mu\nu}=m\, \dot X^\mu \dot X^\nu\,.
\eeq
In principle the coefficient $m$ can also depend on $\tau$. This tensor
is understood to be localized at $x^\mu=X^\mu(\tau)$. 

In effect, we are replacing the entire near-zone geometry with a pointlike
object. In the language of effective field theories, we are
`integrating out' the short-wavelength degrees of freedom of the
near-zone, and replacing them with an effective worldline theory of a
point particle. The coefficient $m$ must then be related through a
matching calculation to a parameter of the `microscopic' configuration, which
in this case can only be the horizon size $r_0$. The matching condition
is that the gravitational field of the effective source reproduces that
of the black hole in the far zone. Thanks to the
separation of
scales \eqref{sepscales} the near and
far zones overlap in
\beq\label{overlap}
r_0\ll r\ll R\,
\eeq
so we can match the fields there. In this region, the near-zone
Schwarzschild solution is in a weak-field regime and can be linearized
around Minkowski spacetime. On the other hand, the background curvature of the
far-zone geometry can be neglected in \eqref{overlap}, so the far field
is the linear perturbation of Minkowski spacetime sourced by
\eqref{efftmunu}. It is now clear that these two fields are the same if
$m$ is equal to the ADM mass of the Schwarzschild solution with horizon
radius $r_0$, so\footnote{Note that $m=\dot X^\mu \dot X^\nu
T_{\mu\nu}=T_{\tau\tau}$ is proportional, but not necessarily equal, to
the mass measured at asymptotic infinity in the background. The relation
is given by the redshift between the particle's
proper time and the asymptotic time of the background.}
\beq
m=\frac{(D-2)\Omega_{D-2}}{16\pi G}r_0^{D-3}\,,
\eeq
where $\Omega_{D-2}$ is the volume of the unit $(D-2)$-sphere.
This result is the first step in the method of ``matched asymptotic
expansions'' \cite{Poisson:2003nc}. One can now proceed to solve the Einstein equations in a
perturbative manner, first in the far-zone, including the backreaction
from the particle with suitable asymptotic conditions, and then in
near-zone, where horizon regularity is imposed. At each step, the
solution in one zone provides boundary conditions for the equations to
be solved in the other zone, through the matching of fields in the overlap region.
Corrections involve higher derivatives of the velocity and of $r_0$.

In this matching construction, a subset of the Einstein equations can
be written as equations for $r_0(\tau)$ and $X^\mu(\tau)$. Their
derivation is a well-understood but technically complicated procedure.
Fortunately, if we are only interested in the leading order equations we
can use a shortcut to them by applying a main guiding principle of
effective theories: symmetry and conservation principles.
In this case the principle is general covariance, which imposes that
\beq\label{geod0}
\oln_\mu T^{\mu\nu}=0\,.
\eeq
This is indeed naturally required for a source of the gravitational
field in the far zone. Put a bit more fancifully, this equation ensures
the consistency of the coupling between short- and long-wavelength
degrees of freedom.

The overline in the covariant
derivative in \eqref{geod0} takes into
account that this makes sense only
when projected along the effective particle trajectory,
\beq
\oln_\mu=-\dot X_\mu \dot X^\nu\nabla_\nu\,.
\eeq
Nevertheless the equation \eqref{geod0} has components both in
directions orthogonal and parallel to the particle's velocity
\beqa\label{geodencon}
(g_{\rho\nu}+\dot X_\rho \dot X_\nu)\oln_\mu T^{\mu\nu}=0\quad \Rightarrow&\quad  &m a^\mu=0\,,\\
\dot X_\nu\oln_\mu T^{\mu\nu}=0\quad \Rightarrow&\quad  &\partial_\tau m(\tau)=0\,,\label{dtm}
\eeqa
with $a^\mu=D_\tau \dot X^\mu=\dot X^\nu\nabla_\nu \dot X^\mu$ being the
effective particle's acceleration. 
The first of these equations is the
geodesic equation that determines the trajectory of a test
particle.\footnote{There is a long history of deriving the geodesic
equation for a small particle (not necessarily a black hole) from the
Einstein field equations, see
\cite{Gralla:2008fg} for a recent rigorous version. This can be taken as
confirmation of our generic symmetry argument.}
The second equation implies that $m$ is a constant along the
trajectory. 

Geodesic motion is so familiar that the effective theory for a black
hole becomes of real interest only when it includes the corrections to
the leading order equations \cite{Poisson:2003nc}. In contrast, we will
see that for black $p$-branes with $p>0$ the effective theory already
yields non-trivial results at leading order. The method of
matched asymptotic expansions (or the closely related classical
effective theory of \cite{Goldberger:2004jt}) provides the conceptual
backdrop to the blackfold approach, but at the practical level we will
remain at the leading order approximation where the black brane is a
`test brane' in a background spacetime.

\section{Effective blackfold theory}

\subsection{Collective field variables}

Our aim is to extend the effective worldline theory of black holes to a
worldvolume theory that describes
the collective dynamics of
a black $p$-brane. 

The geometry of a flat, static black $p$-brane in $D$
spacetime dimensions is 
\beq\label{pbrane}
ds^2=
-\left(1-\frac{r_0^n}{r^n}\right)dt^2+
\sum_{i=1}^p ({dz^i})^2+\frac{dr^2}{1-\frac{r_0^n}{r^n}}+r^2 d\Omega^2_{n+1}
\,.
\eeq
Like in the previous example, the parameters of this solution include the
`horizon thickness' $r_0$ and 
the $D-p-1$ coordinates that parametrize the position of the brane in
directions transverse to the worldvolume, which we denote
collectively by $X^\perp$ (making them explicit in the metric is
possible but cumbersome). But
now we must also include the possibility of a velocity $u^i$ along the
worldvolume of the brane.
A covariant form of the boosted black brane metric is obtained by first introducing the
coordinates $\sigma^a=(t,z^i)$ that span the brane
worldvolume with Minkowski metric $\eta_{ab}$, and a velocity $u^a$
such that $u^a u^b\eta_{ab}=-1$. Then
\beq\label{boostpbrane}
ds^2=
\left(\eta_{ab}+\frac{r_0^n}{r^n}u_a
u_b\right)d\sigma^a d\sigma^b+\frac{dr^2}{1-\frac{r_0^n}{r^n}}+r^2 d\Omega^2_{n+1}
\,.
\eeq

Constant shifts of $r_0$, of $u^i$, and of
$X^\perp$ still give solutions to the Einstein
equations. In total, these are $D$ zero-modes that yield $D$ collective
coordinates of the black brane. The long-wavelength
effective theory describes fluctuations in which these variables change
slowly on the worldvolume
$\mathcal{W}_{p+1}$, over a large
length scale $R\gg r_0$.
Typically $R$ is set by the smallest extrinsic curvature
radius of $\mathcal{W}_{p+1}$, or by the gradient of $\ln r_0$ along the
worldvolume. Background curvatures may also be present but they are
usually already accounted for by the extrinsic curvature they induce on
$\mathcal{W}_{p+1}$.

With this variation the worldvolume metric deviates from the Minkowski
metric $\eta_{ab}$, and the near-zone geometry is of the form
\beq\label{nearpbrane}
ds^2
=
\left(\gamma_{ab}\left(X^\mu(\sigma)\right)+\frac{r_0^n(\sigma)}{r^n}u_a(\sigma)
u_b(\sigma)\right)d\sigma^a d\sigma^b+
\frac{dr^2}{1-\frac{r_0^n(\sigma)}{r^n}}+r^2 d\Omega^2_{n+1}+\dots
\eeq
where the dependence of $\gamma_{ab}$ on the transverse coordinates
gives rise to extrinsic curvature of the worldvolume, and the
 dots indicate that additional terms, of order
$O(r_0/R)$, are required for this to be a solution to Einstein's
equations.\footnote{This is very similar to the long-wavelength
perturbation of
anti-deSitter black branes studied in \cite{chap:fluidgrav}.} 

When $r\gg r_0$ this metric must match the geometry of the far-zone background
with metric $g_{\mu\nu}$, in the region $r\ll R$ around the worldvolume
of an infinitely thin brane at $x^\mu=X^\mu(\sigma)$. Thus we
identify $\gamma_{ab}$ with the metric induced on the effective brane
worldvolume
\beq\label{gammaalbe}
\gamma_{ab}=g_{\mu\nu}\partial_a X^\mu \partial_b X^\nu\,.
\eeq
Again, we are replacing the near-zone geometry with an infinitely thin
$p$-brane, with worldvolume $\mathcal W_{p+1}$, embedded in the background
geometry.

\subsection{Effective stress-energy tensor}
\label{sec:efftensor}

The stress-energy tensor of the black brane is, like the mass $m$ of the black
hole in the previous example, computed in the
overlap region $r_0\ll r\ll R$ where the deviations away from Minkowski
spacetime are small. It can be obtained as a generalization of the ADM
mass, or equivalently, from the Brown-York quasilocal stress-energy
tensor \cite{Brown:1992br}. This is
computed on a timelike surface 
with induced metric $\tilde h_{\mu\nu}$ (not to be
confused with $h_{\mu\nu}$ below) and extrinsic curvature $\Theta_{\mu\nu}$, as
\beq\label{TBY}
T_{\mu\nu}^{\mathit(ql)}=\frac{1}{8\pi G}\left(\Theta_{\mu\nu}-\tilde
h_{\mu\nu}\Theta\right)\,.
\eeq
When measured at constant $r\gg r_0$, the divergent
contributions to this tensor, which grow with $r$, can be subtracted in any of the
conventional ways; for instance, the method of background subtraction from Minkowski
space is enough for our purposes. Then, equivalently, this is the
stress-energy tensor of a domain wall that encloses empty space and creates a field
outside it equal to that of the black brane. This interpretation fits
well with the idea that we replace the black brane with an
effective source.

The surface at large constant $r$ where the quasilocal
$T_{\mu\nu}^{\mathit(ql)}$ is computed has
geometry $\mathbb{R}^{1,p}\times S^{n+1}$. We integrate it
over the
sphere to obtain the stress-energy tensor of the black $p$-brane
\beq
T_{ab}=\int_{S^{n+1}}T_{ab}^{\mathit(ql)}\,,
\eeq
with components along the worldvolume directions. 

For the boosted black $p$-brane \eqref{boostpbrane} the result of this calculation is
\beq\label{blackTab}
T^{ab}=\frac{\Omega_{(n+1)}}{16\pi G}r_0^n\left( n u^a u^b- \eta^{ab}
\right)\,.
\eeq
This is the stress-energy tensor of an isotropic perfect fluid,
\beq\label{perfluid}
T^{ab}=(\vep+P) u^a u^b +P \eta^{ab}\,,
\eeq
where the energy density $\vep$ and pressure $P$ are
\beq\label{blackepsP}
\vep =\frac{\Omega_{(n+1)}}{16\pi G}(n+1)r_0^n\,,\qquad
P=-\frac{\Omega_{(n+1)}}{16\pi G}r_0^n\,.
\eeq
In the rest frame of the fluid, and at any given point on the
worldvolume, the Bekenstein-Hawking
identification between horizon area and
entropy applies to the black hole obtained by compactifying the $p$
directions along the brane. Thus we identify an entropy density from the
horizon area density of \eqref{pbrane},
\beq\label{locs}
s=\frac{\Omega_{(n+1)}}{4 G} r_0^{n+1}\,.
\eeq
Locally, we also have the conventional relation between surface gravity and temperature 
\beq\label{locT}
\mathcal{T}=\frac{n}{4\pi r_0}\,,
\eeq
in such a way that the first law of black hole thermodynamics applies in
the local form
\beq\label{firstlaw}
d\vep =\mathcal{T}ds\,.
\eeq
In addition, the Euler-Gibbs-Duhem relation
\beq\label{duhem}
\vep+P =\mathcal{T}s\,
\eeq
is verified. 

After introducing a slow variation of the collective coordinates, the
stress-energy tensor becomes
\beq\label{blackTab2}
T^{ab}(\sigma)=\frac{\Omega_{(n+1)}}{16\pi G}r_0^n(\sigma)\left( n
u^a(\sigma) u^b(\sigma)- \gamma^{ab}(\sigma)
\right)+\dots
\eeq
where the dots stand for terms with gradients of $\ln r_0$, $u^a$, and
$\gamma_{ab}$, responsible for dissipative effects that
we are taking to be small. We neglect them for
now, but will return to some of them in section~\ref{sec:GLinst}. 

\subsection{Blackfold dynamics}

In order to obtain the equations for the collective variables we need to
recall a few notions about the geometry of worldvolume embeddings. More
details and proofs are provided in the appendix.

\subsubsection{Worldvolume geometry}

The worldvolume $\mathcal{W}_{p+1}$ is embedded in a background with
metric $g_{\mu\nu}$, and its induced metric is \eqref{gammaalbe}.
Indices $\mu\,,\nu$ are raised and lowered with $g_{\mu\nu}$, and $a,b$
with $\gamma_{ab}$. The
first fundamental form of the submanifold
\beq\label{defhmn}
h^{\mu\nu}=\partial_a X^\mu \partial_b X^\nu \gamma^{ab}\,
\eeq
acts as a projector onto $\mathcal{W}_{p+1}$ (for a worldline,
$h^{\mu\nu}=-\dot X^\mu\dot X^\nu$), and the tensor 
\beq\label{orthoproj}
\perp_{\mu\nu}=g_{\mu\nu}-h_{\mu\nu}
\eeq
projects along directions
orthogonal to it. 

Background tensors ${t^{\mu\dots}}_{\nu\dots}$ can be converted into
worldvolume tensors
${t^{a\dots}}_{b\dots}$ and viceversa using the
pull-back map $\partial_a
X^\mu$, a relevant example being the stress-energy tensor
\beq
T^{\mu\nu}=\partial_a X^\mu\partial_b X^\nu\,  T^{ab}\,.
\eeq 

The covariant differentiation of these background
tensors is well defined only along
tangential directions, which we denote by an overline,
\beq
\oln_\mu={h_\mu}^\nu\nabla_\nu\,.
\eeq 
The divergence of the
stress-energy tensor, projected parallel to $\mathcal{W}_{p+1}$,
satisfies (see \eqref{divs})
\beq\label{divT}
{h^\rho}_\nu \oln_\mu T^{\mu\nu}=\partial_b X^\rho D_a T^{ab}\,.
\eeq

The \textit{extrinsic curvature tensor}
\beq\label{Kext}
{K_{\mu\nu}}^\rho={h_\mu}^\sigma \oln_\nu {h_\sigma}^\rho
=-{h_\mu}^\sigma \oln_\nu {\perp_\sigma}^\rho
\eeq
is tangent to $\mathcal{W}_{p+1}$ along its (symmetric) lower indices
$\mu$, $\nu$, and orthogonal to $\mathcal{W}_{p+1}$ along $\rho$. Its
trace is
the \textit{mean curvature vector}
\beq
K^\rho=h^{\mu\nu}{K_{\mu\nu}}^\rho =\oln_\mu h^{\mu\rho}\,.
\eeq

A useful result is that for any two vectors $s^\mu$ and $t^\mu$ tangent to
$\mathcal{W}_{p+1}$,
\beq\label{stK}
s^\mu t^\nu {K_{\mu\nu}}^\rho={\perp^\rho}_\mu \nabla_{s} t^\mu={\perp^\rho}_\mu \nabla_{t} s^\mu\,.
\eeq

\subsubsection{Blackfold equations}

The classical dynamics of a generic brane has been studied by Carter in
\cite{Carter:2000wv}. The equations are formulated in terms of a
stress-energy tensor supported on, and tangent to, the $p+1$-dimensional brane worldvolume
$\mathcal{W}_{p+1}$,
\beq
{\perp^\rho}_\mu T^{\mu\nu}=0\,.
\eeq

As in the example of $p=0$, general covariance implies that
the stress-energy tensor must obey the
equations
\beq\label{nablatmunu}
\oln_\mu T^{\mu\rho}=0\,.
\eeq
This is a consequence of the underlying conservative dynamics of the
full General Relativity theory, but the effective worldvolume theory may be
dissipative.

The divergence in \eqref{nablatmunu} can be written as
\beqa
\oln_\mu T^{\mu\rho}&=&\oln_\mu(T^{\mu\nu}{h_\nu}^\rho)=
T^{\mu\nu}\oln_\mu{h_\nu}^\rho+{h_\nu}^\rho\oln_\mu T^{\mu\nu}\nonumber\\
&=& T^{\mu\nu} {K_{\mu\nu}}^\rho +\partial_b X^\rho D_a T^{ab}\,,
\eeqa
where in the last line we used \eqref{divT} and \eqref{Kext}.
Thus the $D$ equations \eqref{nablatmunu} separate into $D-p-1$ equations
in directions orthogonal to $\mathcal{W}_{p+1}$ and $p+1$ equations parallel
to $\mathcal{W}_{p+1}$,
\beqa
T^{\mu\nu} {K_{\mu\nu}}^\rho&=&0 \qquad 
\mathrm{(extrinsic\ equations)}\,,\label{extreqs}\\
D_a T^{ab}&=&0\qquad \mathrm{(intrinsic\ equations)}\,.\label{intreqs}
\eeqa

The extrinsic equations can be regarded as a generalization to
$p$-branes of the geodesic equation \eqref{geodencon} (where the
acceleration is the extrinsic curvature of the worldline,
$a^\rho=-K^\rho$). In other words, this is the generalization of Newton's
equation ``mass$\times$acceleration$=0$" to relativistic extended
objects. The second set of equations, \eqref{intreqs}, express
energy-momentum conservation on the worldvolume. For a black hole this was a
rather trivial equation, but for a $p$-brane we get all the
complexity of the hydrodynamics of a perfect
fluid. 

If we insert the stress-energy tensor of the black brane
\eqref{blackTab} and use \eqref{stK}, the extrinsic equations
\eqref{extreqs} become
\beq\label{bfeq1}
K^\rho=n{\perp^\rho}_\mu \dot u^\mu\,,
\eeq
and the intrinsic equations \eqref{intreqs},
\beq\label{bfeq2}
\dot u_a+\frac{1}{n+1}u_a D_b u^b =\partial_a\ln r_0\,.
\eeq
Here $\dot u^\mu=u^\nu\nabla_\nu u^\mu$ and $\dot u^b=u^c D_c u^b$. 

Eqs.~\eqref{bfeq1} and \eqref{bfeq2} are the \textit{blackfold
equations}, a set of $D$ equations that describe the dynamics of the $D$ collective
variables of a neutral black
brane, in the approximation where we neglect  its backreaction on the
background (`test brane') as well as the dissipative effects on its
worldvolume. 

\subsection{Blackfold boundaries}
\label{sec:bfbdry}

The worldvolume of the black $p$-brane may have boundaries specified by
a function
$f(\sigma^a)$ such that
$f|_{\partial\mathcal{W}_{p+1}}=0$. If the effective fluid remains
within these boundaries, the velocity must remain parallel
to them,
\beq
\left.u^a\partial_a f\right|_{\partial\mathcal{W}_{p+1}}=0\,.
\eeq

If the boundary is `free', i.e., there is
no surface tension, then the Euler (force) equations
for a generic perfect fluid require that the pressure vanishes at
the boundary.
For the black brane this is
\beq\label{bdryr0}
r_0\big\vert_{\partial\mathcal{W}_{p+1}}= 0\,.
\eeq
Geometrically, this means that the horizon must approach zero size
at the boundary, so the horizon closes off at the edge of the
blackfold. 

\subsection{Blackfold as a black hole}

The blackfold construction puts, on any point in the worldvolume
$\mathcal{W}_{p+1}$, a (small) transverse sphere $s^{n+1}$ with
Schwarzschild radius $r_0(\sigma)$. 
If $\mathcal{B}_p$ is a spatial section of $\mathcal{W}_{p+1}$, then the
geometry of the horizon is the product of $\mathcal{B}_p$ and $s^{n+1}$
--- the product is warped since the radius $r_0(\sigma)$ of $s^{n+1}$
varies along $\mathcal{B}_p$. If $r_0$ is non-zero everywhere on
$\mathcal{B}_p$ then the $s^{n+1}$ are trivially fibered on
$\mathcal{B}_p$ and the horizon topology is the product topology of $\mathcal{B}_p$
 and the sphere. 

The regularity of
this horizon in the perturbative expansion, in which it is distorted by
long-wavelength fluctuations, is believed to be satisfied when the
blackfold equations, which incorporate local thermodynamic equilibrium,
are satisfied. A complete proof is still lacking, but refs.~\cite{Emparan:2007wm} and
\cite{Camps:2010br} provide evidence that this is true, respectively, for extrinsic
and intrinsic deformations.

As we have seen, at boundaries of $\mathcal{B}_p$ the size of $s^{n+1}$
vanishes, so the horizon topology will be different. The regularity of
the horizon at these boundaries is not fully understood yet and appears
to depend on the specific type of
boundary. We will return to this issue later.

\section{Stationary blackfolds, action principle and thermodynamics}

\label{sec:stationary}

Equilibrium configurations for blackfolds that remain stationary in time are of
particular interest since they correspond to stationary
black holes. Requiring stationarity allows to solve explicitly the intrinsic
equations for the thickness
$r_0$ and velocity $u^a$, so one is left only with the extrinsic equations
for the worldvolume embedding $X^\mu(\sigma)$. 

\subsection{Solution to the intrinsic equations}

For a fluid configuration to be stationary, dissipative effects must be
absent. In general, this requires that the shear and expansion of its velocity field
$u$ vanish. One can then prove, using the fluid equations,
that $u$ must be proportional to a (worldvolume) Killing field
$k=k^a\partial_a$. That is,
\beq\label{uzeta}
u=\frac{k}{|k|}\,,\qquad |k|=\sqrt{-\gamma_{ab}k^a k^b}
\eeq
where  $k$
satisfies the worldvolume
Killing equation $D_{(a}k_{b)}=0$.
Actually, we will assume that this Killing vector on the worldvolume is the 
pull-back of a timelike Killing vector $k^\mu \partial_\mu$ in the
background, 
\beq\label{Killbck}
\nabla_{(\mu}k_{\nu)}=0\,,
\eeq
such that $k_a=\partial_a X^\mu k_\mu$. The
existence of this timelike Killing vector field is in
fact a necessary assumption if we intend to describe stationary
black holes.

The Killing equation \eqref{Killbck} implies
\beq
\nabla_{(\mu}u_{\nu)}=-u_{(\mu}\nabla_{\nu)}\ln|k|\,,
\eeq
so the acceleration is
\beq\label{unuz}
\dot u^\mu=\partial^\mu \ln |k|\,.
\eeq
Since the expansion of $u$ vanishes, the intrinsic equation
\eqref{bfeq2}  becomes
\beq
\partial_a \ln |k|=\partial_a\ln r_0\,
\eeq
so
\beq\label{r0zeta}
\frac{r_0}{|k|}=\mathrm{constant}\,.
\eeq
Expressed in terms of the local temperature $\mathcal T$, \eqref{locT},
this equation says that the 
worldvolume variation of the temperature is dictated by the local redshift factor
$|k|^{- 1}$,
\beq\label{redshiftT}
\mathcal T(\sigma)=\frac{T}{|k|}\,.
\eeq
This result can also be derived for a general fluid using the equations of
fluid dynamics.
The integration constant $T$ can be interpreted, using the thermodynamic
first law that we derive below, as the global temperature of the black
hole.
Equation \eqref{r0zeta} can be read as saying that the thickness 
\beq\label{r0k}
r_0(\sigma)=\frac{n}{4\pi T} |k|
\eeq
adjusts its value along the
worldvolume so that $T$ is a constant.

\subsection{Extrinsic equations and action for stationary blackfolds}
\label{sec:extreqs}

With the intrinsic solutions \eqref{unuz} and
\eqref{r0zeta}, the
extrinsic equations \eqref{bfeq1} reduce to
\beqa\label{extrinsic}
K^\rho
&=&n\perp^{\rho\mu}\partial_\mu\ln r_0\nonumber\\
&=&\perp^{\rho\mu}\partial_\mu\ln(-P)\,.
\eeqa

Using eq.~\eqref{varI} from the appendix, this equation can be
equivalently found by varying, under deformations of the brane
embedding, the action
\beq
I=\int_{\mathcal{W}_{p+1}}d^{p+1}\sigma\sqrt{-h}\,P\,.
\label{stataction}
\eeq
This action, whose derivation actually need not assume any specific
fluid equation of state, is a familiar one for 
branes with constant tension $-P$, whose worldvolume must be a minimal
hypersurface so $K^\rho=0$ (an example are Dirac-Born-Infeld branes with
zero gauge fields on their worldvolume). More generally, this is the
action of a perfect fluid on $\mathcal{W}_{p+1}$.

Assume now that the background spacetime has a timelike Killing
vector $\xi$, canonically normalized to generate unit time translations
at asymptotic
infinity, and whose norm on the worldvolume is
\beq
\left.-\xi^2\right|_{\mathcal W_{p+1}}=R_0^2(\sigma)\,.
\eeq
Let us further assume that $\xi$ is hypersurface-orthogonal, so we can
foliate the blackfold in spacelike slices $\mathcal{B}_p$ normal to $\xi$.
The unit normal to $\mathcal{B}_p$ is
\beq\label{redshift}
n^a =\frac1{R_0}\xi^a\,.
\eeq
$R_0$ measures the local gravitational redshift between worldvolume time and
asymptotic time. Integrations over $\mathcal W_{p+1}$ reduce, over an
interval $\Delta t$ of the Killing time generated by $\xi$, to integrals
over $\mathcal{B}_p$ with measure $dV_{(p)}$, so
\beq\label{stataction2}
I=
\Delta t\int_{\mathcal{B}_p}dV_{(p)}\,R_0P\,.
\eeq
 Using
\eqref{r0k} in \eqref{blackepsP} we get an expression for the action in
terms of $k$ that is very practical for
deriving the extrinsic equations in explicit cases,
\beq
\tilde I[X^\mu(\sigma)]=\int_{\mathcal{B}_p}dV_{(p)}\,R_0|k|^n\,.
\label{action}
\eeq
The tilde distinguishes it from \eqref{stataction2}, since we have
removed an overall constant factor (including a sign) that is irrelevant
for obtaining the equations.

\subsection{Mass, angular momentum, entropy, and thermodynamics}

Let $k$ be given by a linear
combination of orthogonal commuting Killing vectors of the background
spacetime,
\beq\label{kxichi}
k=\xi+\sum_i\Omega_i\chi_i,
\eeq
where $\xi$ is the generator of time-translations that
we introduced above, and $\chi_i$ are
generators of angular rotations in the background, normalized
such that their orbits have periods $2\pi$. Then $\Omega_i$ are the
angular velocities of
the blackfold along these directions.

The mass and angular momenta of the blackfold are now given by the integrals of the
energy and momentum densities over $\mathcal B_p$,
\beq
M=\int_{\mathcal B_p}dV_{(p)}\,T_{ab}n^a\xi^b\,,\qquad
J_i=-\int_{\mathcal B_p}dV_{(p)}\,T_{ab}n^a\chi_i^b\,.
\label{bfMJ}\eeq
The total entropy is deduced from the entropy current $s^a=s(\sigma)u^a$,
\beq
S=-\int_{\mathcal B_p}dV_{(p)}\,s_a n^a=
\int_{\mathcal B_p}dV_{(p)}\,\frac{R_0}{|k|} s(\sigma)\,.
\label{bfS}\eeq

Let us now express the action \eqref{stataction2} in terms of these
quantities. Contracting the stress-energy tensor
\eqref{perfluid} with $n_a k_b$, then using \eqref{duhem} and \eqref{redshiftT}, we find
\beq\label{manip}
T_{ab}\,n^a\left(\xi^b+\sum_i\Omega_i \chi_i^b\right)
+T su^an_a=n^a k_a P=-R_0 P\,,
\eeq
so, integrating over $\mathcal B_p$,
\beq\label{Ithermo}
I=-\Delta t\left( M-TS-\sum_i \Omega_i J_i \right)\,.
\eeq
This is an action in real Lorentzian time, but since we are dealing with
time-independent configurations we can rotate to Euclidean
time
with periodicity $1/T$ and recover the relation between the
Euclidean action and the
thermodynamic grand-canonical potential, $I_E=W[T,\Omega_i]/T$. 

The identity
\eqref{Ithermo} holds for any embedding, not necessarily a solution to
the extrinsic equations, so if we regard $M$, $J_i$ and $S$ as
functionals of the $X^\mu(\sigma)$
and
consider variations where
$T$ and $\Omega_i$ are held constant, we have
\beq\label{bhfirstlaw}
\frac{\delta I[X^\mu]}{\delta X^\mu}=0\quad \Leftrightarrow\quad
\frac{\delta M}{\delta X^\mu}=T
\frac{\delta S}{\delta X^\mu}+\sum_i\Omega_i\frac{\delta J_i}{\delta X^\mu}\,.
\eeq
Therefore,
the extrinsic equations are equivalent to the requirement that the first
law of black hole thermodynamics holds for variations of the embedding.

Eq.~\eqref{Ithermo}, Wick-rotated to $I_E[X^\mu]$, is therefore an effective worldvolume
action that approximates, to leading order in $r_0/R$, the Euclidean
gravitational action of the black hole. One might have
taken this thermodynamic effective action as the starting point for the
theory of stationary blackfolds, but we have preferred to work with the
equations of motion. These allow to consider situations away from
stationary equilibrium and they also make more explicit the connection with
worldvolume fluid dynamics.

Performing manipulations similar to \eqref{manip} one finds that
\beq\label{presmarr}
(D-3)M-(D-2)\left( TS+\sum_i\Omega_i J_i\right)=
\mbox{\boldmath$\mathcal T_{\text{tot}}$}\,,
\eeq
where
\beq\label{tenstot}
\mbox{\boldmath$\mathcal T_{\text{tot}}$}=-\int_{\mathcal B_{p}}\!dV_{(p)}\, 
R_0\left(\gamma^{ab}+n^an^b\right) T_{ab}
\eeq
is the total tensional energy, obtained by integrating the local tension
 over the blackfold volume.

The Smarr relation for asymptotically flat vacuum black holes in $D$ dimensions \cite{Harmark:2003dg},
\beq\label{smarr}
(D-3)M-(D-2)\left( TS+\sum_i\Omega_i J_i\right)=0\,
\eeq
must be recovered when the extrinsic
equations for equilibrium are satisfied for a blackfold with compact
$\mathcal B_{p}$ in a Minkowski background where
$R_0=1$. Thus, extrinsic equilibrium in
Minkowski backgrounds implies
\beq\label{zerotension}
\mbox{\boldmath$\mathcal T_{\text{tot}}$}=0\,.
\eeq
If the tensional energy did not vanish, it would imply the presence of
sources of tension acting on the blackfold, e.g., in the form of conical or
stronger singularities of the background space.

Another general identity is obtained by noticing that the
blackfold fluid satisfies
\beq
-P =\frac{1}{n}\mathcal T s\,.
\eeq
Upon integration and using \eqref{Ithermo} we get
\beq\label{ITS}
M-TS-\sum_i \Omega_i J_i=\frac1{n}T S\,,
\eeq
or, in terms of the Euclidean action, $I_E=S/n$.

Note that while the thermodynamic form of the action \eqref{Ithermo} and
the Smarr relation \eqref{smarr} are exactly valid for neutral black
holes, eqs.~\eqref{presmarr} and \eqref{ITS} instead hold only to
leading order in the expansion in $r_0/R$.

\subsection{Stationary blackfold boundaries}

Let us now investigate what it means, for a stationary blackfold, that
the thickness vanishes at
its boundary, eq.~\eqref{bdryr0}.

In section~\ref{sec:extreqs} we have introduced the generators of unit
time translations at asymptotic
infinity, $\xi^a$, and on $\mathcal
W_{p+1}$, $n^a$, which are related by the factor $R_0$
that measures the gravitational
redshift between these two locations. On the other hand,
relative to the worldvolume time generated by $n^a$, a fluid element on $\mathcal W_{p+1}$
has a Lorentz-boost gamma factor equal to
\beq\label{naua}
-n^a u_a=\frac1{\sqrt{1-v^2}}
\eeq
where $v$ is the local fluid velocity, 
\beq
v^2=\sum_i v_i^2\,,\qquad
v_i=\frac{\Omega_i|\chi_i|}{R_0}\,.
\eeq

Since the velocity $u^a$ is determined by \eqref{uzeta} and
\eqref{kxichi}, then $\xi^a u_a=\xi^a\xi_a/|k|=-R_0^2/|k|$, and
\beq
-n^a u_a=-\frac1{R_0}\xi^a u_a=\frac{R_0}{|k|}\,.
\eeq
With \eqref{naua}, this implies 
\beq
|k|=R_0\sqrt{1-v^2}\,.
\eeq 

At a blackfold boundary we must have $r_0\to
0$. According to \eqref{r0zeta}, if the blackfold is stationary it must
also be that $|k|\to 0$. There are two possibilities: 
\begin{itemize}
\item[(i)] $v\to 1$: the fluid velocity becomes null at the
boundary. There is some evidence that in this case the full horizon is
smooth: as we will see in section \ref{subsec:MPbf}, there
are blackfolds with this kind of boundary for which we can compare to an exact
black hole solution with a regular horizon.

\item[(ii)] $R_0\to 0$: the blackfold encounters a horizon of the
background, where the gravitational redshift diverges. This boundary 
can be regarded as an intersection of horizons, and the evidence from known exact
solutions indicates that the intersection
point, i.e., the blackfold boundary, is singular.

\end{itemize}

Nevertheless, the evidence for these two behaviours is largely
circumstantial and it would be desirable to have a better understanding
of horizon regularity at blackfold boundaries.

\subsection{Ultraspinning behaviour}

Let us assume that all length scales along $\mathcal B_{p}$ are of the
same order $\sim R$ and that there are no large redshifts, of
gravitational or Lorentz type, over most of the blackfold --- this is
naturally satisfied since the redshifts become large only near the
boundaries.
Then (setting temporarily $G=1$) eqs.~\eqref{bfMJ} and \eqref{bfS} generically give
\beq
M\sim R^p r_0^n\,,\qquad J\sim R^{p+1}r_0^n\,,\qquad
S\sim R^p r_0^{n+1}\,.\label{scalings}
\eeq
This implies
\beq\label{JonM}
\frac{J}{M}\sim R\,,
\eeq
so that neutral blackfolds are always in an ultraspinning regime, in which the
angular momentum for fixed mass is very large. More precisely, in a neutral blackfold the length
scale of angular-momentum
\eqref{JonM} is always much larger than the
length scale of the mass $M^{1/(D-3)}$,
\beq
\frac{J/M}{M^{1/(D-3)}}\sim \left(\frac{r_0}{R}\right)^\frac{D-p-3}{D-3}\ll 1\,.
\eeq

The entropy in \eqref{scalings} scales like
\beq
S(M,J)\sim J^{-\frac{p}{D-p-3}}M^{\frac{D-2}{D-p-3}}\,,
\eeq
so, in dimension $D$ and for fixed mass, the most entropic solution
for a given number of large angular momenta $J$ is attained by
blackfolds with the smallest $p$. The intuitive reason is that, for a
given mass, the horizon is thicker if $p$ is smaller --- the horizon
spreads out less. A thicker horizon has lower
temperature, and since $TS\sim M$, the entropy is higher.
In section \ref{subsec:helical} we will find that there is always a
black 1-fold for any number of angular
momenta, which therefore maximizes
the entropy.

\section{Examples of blackfold solutions}
\label{sec:bhsbfs}

This formalism can be applied easily to the explicit construction of
stationary black holes. Besides finding new solutions, we will show that
the blackfold method correctly recovers the limit in which known exact
solutions become similar to black branes --- these are the ultraspinning
regime of Myers-Perry black holes and the very-thin limit of the
five-dimensional black ring, which provide non-trivial checks on the
method.

\subsection{Myers-Perry black hole as a blackfold disk}
\label{subsec:MPbf}

Myers-Perry black holes have ultraspinning regimes where the geometry
near the rotation axis approaches that of a black brane spread along the
rotation plane \cite{Emparan:2003sy}. This suggests that these regimes may be
reproduced by blackfold constructions, and indeed they are, in a rather
non-trivial manner. Instead of studying the most
general construction (see \cite{Emparan:2009vd}), we will illustrate it in the case of
the six-dimensional black hole rotating along a single plane, which
already exhibits all the relevant features.

Take $D=6$ Minkowski spacetime as a background and a black 2-fold that
extends along a plane within it (so $n=1$). The extrinsic equations
are
trivially solved, and we can restrict the analysis to the blackfold
plane with polar coordinates $(r,\phi)$ and metric
\beq
ds^2=-dt^2+dr^2+r^2d\phi^2\,.
\eeq
We set the blackfold in rotation along $\phi$.
Stationarity requires that the fluid rotates rigidly, eqs.~\eqref{uzeta}
and \eqref{kxichi} with
$\xi=\partial/\partial t$,
$\chi=\partial/\partial\phi$ and angular velocity $\Omega$, so the velocity is
\beq
u=\frac{1}{\sqrt{1-\Omega^2r^2}}\left(\frac{\partial}{\partial t}+\Omega
\frac{\partial}{\partial\phi}\right)\,.
\eeq
The intrinsic equations are solved by appropriately redshifting
the temperature, 
which determines the horizon thickness as in \eqref{r0k},
\beq\label{r0disk}
r_0(r)=\frac{1}{4\pi T}\sqrt{1-\Omega^2r^2}\,.
\eeq
This implies that the extent of the blackfold along the
rotation plane is limited to
\beq
0\leq r\leq \Omega^{-1}\,.
\eeq
Recall that according to eq.~\eqref{bdryr0} the condition
$r_0(\Omega^{-1})=0$ specifies a boundary of the blackfold. The
effective fluid velocity becomes lightlike there, and would be
superluminal if we tried to go beyond this radius. Therefore the
blackfold worldvolume is a
disk. 

It will be convenient to introduce two geometric parameters in
place of $T$ and $\Omega$: the disk radius $a$, and the
blackfold thickness $r_+$ at the rotation axis,
\beq
a=\Omega^{-1}\,,\qquad r_+=r_0(0)=\frac{1}{4\pi T}\,.
\eeq

What is the topology of the horizon of this blackfold? The
disk is fibered at
each point with a sphere $S^2$ of radius $r_0(r)$ that shrinks to zero at the
disk's edge. Topologically, this is $S^4$, i.e., the same as the
topology of the horizon of the Myers-Perry black hole.

The mass and angular momentum of the blackfold are obtained from its
energy and momentum densities,
\beqa
T_{ab}n^a\xi^b&=&T_{tt}=\frac{r_+}{4G}\frac{2-r^2/a^2}{\sqrt{1-
r^2/a^2}}\,,\\
-T_{ab}n^a\chi^b&=&T_{t\phi}=\frac{r_+}{4G}\frac{r^2/a}{\sqrt{1-
r^2/a^2}}
\eeqa
(since $n^a=\xi^a$). Then
\beq\label{Mbf}
M=\int_{0}^{2\pi}d\phi\int_{0}^{a} dr\, r\, T_{tt}=
\frac{2\pi}{3G}r_+ a^{2}\,,
\eeq
\beq\label{Jbf}
J=\int_{0}^{2\pi}d\phi\int_{0}^{a} dr\, r\, T_{t\phi}=
\frac{\pi}{3G}r_+ a^{3}\,.
\eeq
The entropy-density current $s^a=s\,u^a $ gives
\beq
-s^a n_a =\frac{\pi}{G}r_+^2\sqrt{1-r^2/a^2}
\eeq
which integrates to
\beq\label{Sbf}
S =-\int_{0}^{2\pi}d\phi\int_{0}^{a}dr\;r\; s^a n_a=\frac{2\pi^2}{3G}r_+^2 a^2\,.
\eeq

Let us now compare to the ultraspinning limit of the exact Myers-Perry
black hole. This solution is specified by a mass
parameter $\mu$ and a rotation parameter $a$. The horizon radius $r_+$ is
obtained as the largest real root of
\beq\label{rproot}
\mu=(r_+^2+a^2)r_+\,.
\eeq
In terms of these, the exact mass, spin and entropy are
\beq\label{exactMJS}
M=\frac{2\pi}{3G}\,\mu\,,\qquad
J=\frac12 aM\,,\qquad S=\frac{2\pi^2}{3G}\,r_+\mu\,.
\eeq
The ultraspinning regime of $J\to\infty$ with fixed $M$ is obtained as
$a\to\infty$. In this limit eq.~\eqref{rproot} becomes
\beq
\mu\to a^2r_+\,,
\eeq
and the physical quantities \eqref{exactMJS} become precisely the same as \eqref{Mbf},
\eqref{Jbf}, \eqref{Sbf}, after identifying the parameters
$r_+$ and $a$ in both sides. 

This identification of parameters is in fact
geometrically meaningful. For an ultraspinning black hole the radii of the horizon
in directions transverse and parallel to the rotation plane are \cite{Emparan:2003sy}
\beq
r_\perp^{MP}\to r_+\cos\theta\,,\qquad r_\|^{MP} \to  a\,,
\eeq
where $\theta$ is the polar angle on the horizon. 
For the blackfold disk, we have already seen that the radius in the plane
parallel to the rotation is $r_\|^{bf} = a$.
The square root in \eqref{r0disk} suggests
to introduce a polar angle $\theta=\arcsin(r/a)$ such that the thickness of
the blackfold in directions transverse to the rotation plane is
\beq
r_\perp^{bf}=r_0(r)=r_+\cos\theta\,.
\eeq
Identifying this polar angle with the one in the Myers-Perry solution,
we find a perfect match between both sides. 

Since the horizon of the Myers-Perry solution is smooth, this example
yields evidence of regularity at the boundary of the
blackfold where the fluid velocity becomes lightlike.

\subsection{Curving black strings: black rings}

We now consider stationary black 1-folds (i.e., curved black strings) in
a Minkowski background 
\beq
ds^2=-dt^2+d\mathbf{x}^2_{(D-1)}
\ee
with
stationarity Killing vector $\xi=\partial/\partial t$ so $R_0=1$.

Stationarity requires that the black string lies along a spatial
Killing direction. Then the
simplest way to solve the extrinsic equations is by imposing that the
tensional energy vanishes,
eq.~\eqref{zerotension}. Since the string lies along an isometry, the
integral in \eqref{tenstot} is trivial and so
the integrand, i.e., the
tension measured relative to the frame defined
by $\xi$, must vanish:
\beq\label{tens0}
\mathrm{tension}=-\left(\gamma^{ab}+\xi^a\xi^b\right)T_{ab}=0\,.
\eeq
Let us denote
\beq
-\xi^a u_a=\cosh\sigma
\eeq
so $\sigma$ is the rapidity that parametrizes the relative boost between the fluid
and the background frame of observers along orbits of $\xi$. For a
generic perfect fluid, the zero-tension condition \eqref{tens0} is
\beq
0=\left(\gamma^{ab}+\xi^a\xi^b\right)\Bigl((\vep+P)u_a u_b +P\gamma_{ab}\Bigr)=
\vep\sinh^2\sigma+P\cosh^2\sigma
\eeq
so at equilibrium the fluid velocity $\tanh\sigma$ is fixed to the value
\beq\label{equiltanh}
\tanh^2\sigma=-\frac{P}{\vep}\,.
\eeq
Since $-P/\vep$ is actually the square of the velocity of transverse, elastic
waves along the string, we see that the bending of the string can be
regarded as the effect of supporting a stationary elastic wave along itself.

For the specific black string fluid \eqref{blackepsP}, the condition
\eqref{equiltanh} can be written as
\beq\label{equilsinh}
\sinh^2\sigma=\frac1n\,.
\eeq
When $n=1$, i.e., in $D=5$, this is precisely the value for the boost
that was found in \cite{chap:blackrings} from the exact black ring
solution in the limit where the ring is very thin and long, $r_0/R\to
0$. In this case the black string lies along a circle of radius $R$ on a plane within
$\mathbb{R}^4$.

It is straightforward to calculate \eqref{bfMJ} and \eqref{bfS} and
check that, with this value of the boost, the blackfold construction
reproduces all the physical magnitudes of the exact black ring solution
to leading order in $r_0/R$. Beyond this order,
ref.~\cite{Emparan:2007wm} has computed the first corrections to the
metric, finding again perfect agreement with the perturbative expansion
of the exact solution in $D=5$, and checking that horizon regularity is
preserved in any $D\geq 5$.

\subsection{Helical black rings and minimal rigidity}
\label{subsec:helical}

The previous construction did not specify yet the geometry of the curve
along which the
string lies. However, as we saw, this must be along a spatial isometry of
Euclidean space $\mathbb{R}^{D-1}$. If we are interested in blackfolds
of finite extent, which correspond to asymptotically flat black
objects, this isometry must be compact and therefore generated by rotational
Killing vectors $\partial/\partial \phi_i$. This is, if we write the
subspace of $\mathbb{R}^{D-1}$ in which the embedding of the string is
non-trivial as
\beq\label{subsp}
ds^2=\sum_{i=1}^m \left(dr_i^2 +r_i^2 d\phi_i^2\right)\,,
\eeq
then the string lies along the curve 
\beq\label{curve}
r_i=R_i\,,\qquad \phi_i=n_i \sigma\,,\qquad
0\leq \sigma< 2\pi\,.
\eeq
An upper limit $m\leq\left\lfloor\frac{D-1}{2}\right\rfloor=
\left\lfloor\frac{n+3}{2}\right\rfloor$
is set by the rank of the spatial rotation group in $D=n+4$ spacetime
dimensions. The $D-2m-2$ dimensions of space that are not explicit in
\eqref{subsp} are totally orthogonal to the string and we will ignore
them. They only play a role in providing, together with the $m$
directions $r_i$, the $n+2$ dimensions
orthogonal to the worldsheet in which the horizon of the black string is
`thickened' into a transverse $s^{n+1}$ of radius $r_0$.

The $n_i$ must be integers in order that the curve closes in on itself.
Without loss of generality we assume $n_i \geq 0$. If we want to avoid
multiple covering of the curve then the smallest of the $n_i$ (which
need not be unique), say $n_1$, must be coprime with all the $n_i$. Thus
the set of $n_i$ can be specified by $m$ positive rational numbers
$n_i/n_1\geq 1$.

If all $n_i=1$ we obtain a circular planar ring along an orbit of
$\sum_i\partial_{\phi_i}$ with radius $\sqrt{\sum_i R_i^2}$.
If, instead, at least two $n_i$ are non-zero and not both
equal to one then we obtain {\em helical black
rings}. Together with planar black rings, this
exhausts all
possible stationary black 1-folds in a Minkowski background with a spatially
compact worldsheet. 

The Killing generator
of the worldsheet velocity field is
\beq
k=\frac{\partial}{\partial t}+\sum_i \Omega_i
\frac{\partial}{\partial \phi_i}\,,
\eeq
where the ratios between angular velocities must be rational
\beq\label{ratios}
\left|\frac{\Omega_i}{\Omega_j}\right|=\frac{n_i}{n_j}\quad  \forall
i,j\,,
\eeq
and the equilibrium condition \eqref{equilsinh} fixes
\beq\label{vequil}
\sum_{i=1}^m \Omega_i^2 R_i^2=\frac{1}{n+1}\,.
\eeq

The physical properties (mass, spins, entropy etc) of helical black
rings in the approximation $r_0/R_i\ll 1$ are computed in \cite{Emparan:2009vd}. One
finds that helical black rings are the solutions with the largest
entropy among all blackfolds with given values of the mass and angular
momenta. Among helical black rings, the maximal entropy for a given 
set of values of the angular momenta is achieved by minimizing
the $n_i$, since this makes the ring shorter and hence thicker. For a
single angular momentum, the planar black ring
maximizes the entropy.

\subsubsection{Maximal symmetry breaking and saturation of the rigidity theorem}

A spacetime containing a helical ring
has the isometry generated by
\beq\label{oneu1}
\sum_i n_i \frac{\partial}{\partial \phi_i}
\eeq
along the direction of the string. However, this string breaks in general
other $U(1)$ isometries of the
background, possibly leaving \eqref{oneu1} as the only spatial Killing
vector of the configuration.

In order to prove this point, observe that any additional $U(1)$ symmetry must
leave the curve invariant, i.e., the curve must lie at a fixed point of
the isometry. This is, the curve must be on a point in some plane in
$\mathbb{R}^{D-1}$, so rotations in this plane around the point leave it
invariant. Let us parametrize the most general possible helical curve in
$\mathbb{R}^{D-1}$ as a curve in $\mathbb{C}^m$, with
$m=\left\lfloor\frac{D-1}{2}\right\rfloor$, of the form
\beq
z_i=R_i e^{i n_i \sigma}\,,
\eeq
where possibly some of the
$n_i$ are zero.
In order to find a plane in which rotations leave the curve invariant we must
solve the equation
\beq
\label{aizi}
\sum_{i=1}^m a_i z_i=0
\eeq
with complex $a_i$, for all values of $\sigma$. This equation admits
a non-trivial solution only if some of the $n_i$ are equal to each other
(possibly zero).

Therefore, if (i) the string circles around in all of the
$m=\left\lfloor\frac{D-1}{2}\right\rfloor$ independent rotation planes,
i.e., all the possible $n_i$ are non-zero, and (ii) all the $n_i$'s are
different from each other, then the only spatial Killing vector of the
configuration is \eqref{oneu1}. In this case, we obtain an
asymptotically flat helical black ring with only one spatial $U(1)$
isometry.

The black hole rigidity theorem of \cite{Hollands:2006rj, Moncrief:2008mr} requires
that stationary non-static black holes have at least one such isometry, and
ref.~\cite{Reall:2002bh} had conjectured that black holes exist with not
more than this symmetry. The construction of helical black rings that
rotate in all possible planes and have
exactly one spatial $U(1)$ proves this conjecture in any $D\geq 5$.

\subsection{Odd-spheres}

For our final example, we describe a large family of solutions for
black holes in
$D$-dimensional flat space with horizon topology
\beq \label{oddSaa}
 \Big( \prod_{p_a=\rm odd} S^{p_a} \Big) \times s^{n+1}\,, 
\qquad \sum_{a=1}^\ell
p_{a}=p \,.
\eeq
In this case, the spatial section of the blackfold
worldvolume $\mathcal{B}_p$ is a product of odd-spheres.

\subsubsection{$S^{2k+1}$ blackfolds} 

We consider first a single
odd-sphere $S^{2k+1}$, which contains the black
ring as the particular case $k=0$.
We embed the sphere $S^{2k+1}$
 into a $(2k+2)$-dimensional flat subspace of $\mathbb{R}^{D-1}$
with metric
\beq
\label{oddSab}
d r^2+r^2 d\Omega_{2k+1}^2\,.
\eeq
The unit metric on $S^{2k+1}$ can be written as
\beq
d\Omega_{2k+1}^2=\sum_{i=1}^{k+1}\left( d \mu_i^2+\mu_i^2 d \phi_i^2\right)
\,, \qquad \sum_{i=1}^{k+1} \mu_i^2=1\,,
\eeq
where $\phi_i$ are the angles that parametrize the Cartan subgroup of
the rotation group $SO(2k+2)$.

In the general stationary case we would consider the blackfold embedded as
$r=R(\mu_1,\dots,\mu_k)$. Then the extrinsic equations give a set
of differential equations for $R(\mu_i)$ involving $k$ angular velocities $\Omega_i$.
These are complicated to solve, but they simplify to
algebraic equations in the still non-trivial instance
that the sphere is geometrically round with
constant radius $r=R$, and rotates with the same angular velocity $\Omega$ in all
angles. Then the stationarity Killing vector is
\beq
k=\frac{\partial}{\partial t}+\Omega\sum_{i=1}^{k+1}
\frac{\partial}{\partial \phi_i}\,,\qquad |k|=\sqrt{1-\Omega^2R^2}\,,
\eeq
and the blackfold is homogeneous so the thickness $r_0$ is constant over the worldvolume.

The extrinsic equations for $R$ can be easily (and consistently) obtained
from the stationary blackfold action \eqref{action}
\beq \label{oddSag}
\tilde I[R]= \Omega_{(p)} R^{p}\left( 1-
R^2\Omega^2 \right)^\frac{n}{2}\,, \qquad p=2k+1 \,.
\eeq
This is extremized when
\beq
\label{oddSai} R= \frac{1}{\Omega}\sqrt{\frac{p}{n+p}}\,.
\eeq
Equivalently, this value makes the local tension \eqref{tens0} vanish at
each point. When $p=1$ we recover the result for black rings. Having this
equilibrium radius of round blackfolds, it is
straightforward to compute their physical properties \cite{Emparan:2009vd}.

\subsubsection{Products of odd-spheres}
\label{prododd}

This construction can be easily extended to solutions where
$\mathcal{B}_p$ is a product of round
odd-spheres, each one labeled by an index $a=1,\ldots, \ell$.
Denoting the radius of the $S^{p_a}$ factor ($p_a=$odd) by
$R_a$ we take the angular momenta of the $a$-th sphere to be all
equal to $\Omega^{(a)}$.

We embed the product of $\ell$ odd-spheres
in a flat $(p+\ell)$-dimensional subspace of $\mathbb{R}^{D-1}$ with metric
\beq
\label{oddSba}
\sum_{a=1}^\ell \left( d r_a^2+r^2_a d \Omega^2_{(p_a)}\right) \,,
\qquad
\sum_{a=1}^\ell p_a = p
\eeq
and locate the blackfold at $r_a=R_a$. Note
that given a value of $n=D-p-3\geq 1$, the number $\ell$ of spheres in
the product is limited to $\ell\leq n+2$.

With the spheres being geometrically round, the stationary
blackfold action \eqref{action} reduces to
\begin{equation}
\label{oddSbb}
\tilde I[R_a] = \prod_{b=1}^\ell  \Omega_{(p_b)} R_b^{p_b}
\left( 1  - \sum_{a=1}^\ell \left (R_a \Omega^{(a)}\right )^2
\right)^{n/2}\,,
\end{equation}
whose variation with respect to each of the $R_a$'s gives
the equilibrium conditions
\begin{equation}
\label{oddSbc}
R_a = \frac{1}{\Omega^{(a)}}\sqrt{ \frac{p_a}{n+ p} } \,.
\end{equation}

A simple case is the
$p$-torus, where we set all $p_a =1$. This gives
black holes with horizon topology 
$\mathbb{T}^p \times s^{n+1}$
that rotate simultaneously along all
orthogonal one-cycles of $\mathbb{T}^p$. 

It is easy to see that, like the Myers-Perry black holes and the planar
black rings, these odd-sphere solutions do not break any of the commuting
isometries of the background.

\medskip

To finish this section we note that, except for the case of black
1-folds, our analysis has not been systematic enough to be complete, and
further classes of black holes can be expected in $D\geq 6$. But already
with the ones we have presented, one can easily see that black hole
uniqueness is very badly violated in higher-dimensions.

\section{Gregory-Laflamme instability and black brane viscosity}
\label{sec:GLinst}

The blackfold approach must capture the perturbative dynamics of a black
hole when the perturbation along the horizon has long wavelength $\lambda$,
\beq
\lambda\gg r_0 \,.
\eeq
This includes in particular intrinsic fluctuations of the black
brane in which the worldvolume geometry remains flat but $r_0$ and $u^a$
are allowed to vary. A variation of the thickness of the brane, $\delta
r_0$, is a variation of the pressure and density of the effective fluid.
Then, for small fluctuations we expect to recover sound waves along the
brane. These turn out to be unstable in an interesting way.

\subsection{Sound waves on a black brane}

Sound waves are easily derived for a
generic perfect fluid. Introduce small perturbations in an initial uniform
state at rest,
\beq
\vep\to\vep+\delta\vep\,,\qquad
P\to P+\frac{dP}{d\vep}\delta\vep\,,\qquad
u^a=(1,0\dots)\to (1,\delta u^i)\,.
\eeq
To linear order in the perturbations the intrinsic fluid equations \eqref{intreqs} give
\beq
\left(\partial_t^2-\frac{dP}{d\vep}\partial_i^2 \right)\delta\vep=0\,,
\eeq
so longitudinal, sound-mode oscillations of the fluid propagate with
squared speed
\beq\label{cL}
v_s^2=\frac{dP}{d\vep}\,.
\eeq
Neutral blackfolds have imaginary speed of sound
\beq\label{soundspeed}
v_s^2=-\frac{1}{n+1}\,,
\eeq
which implies that sound waves
along the effective black brane fluid are unstable: under a
density perturbation the fluid evolves to become more and more
inhomogenous. Thus the black brane horizon itself
becomes inhomogeneous, with the brane thickness $r_0$ varying along
the brane as
\beq\label{delr0}
\delta r_0 \sim e^{\Omega t +ik_i z^i}\,, 
\eeq
with 
\beq\label{Omk}
\Omega=\frac{k}{\sqrt{n+1}}\, + O(k^2)\,.
\eeq
($k=\sqrt{k_i k_i}$). Unstable oscillations of the form \eqref{delr0}
are the type of black brane instability discovered by Gregory and
Laflamme (see \cite{chap:GLinstab}). Using the blackfold effective theory, we
have derived it in the regime of long wavelengths, $k r_0\ll 1$. Many
studies of the Gregory-Laflamme instability focus on
the threshold mode, with $\Omega=0$ at $k=k_{GL}\neq 0$, which has `small'
wavelength $2\pi/k_{GL} \sim r_0$ and typically needs numerical work to
determine. The blackfold approach instead reveals that the hydrodynamic
modes, which have vanishing frequency as $k\to 0$, are much simpler to
study.
The slope of the curve $\Omega(k)$ near $k=0$ is exactly determined
using only the equation of state $P(\vep)$ of the unpertubed, static
black brane.

\subsection{Correlated dynamical and thermodynamical stability}

It is conventional in fluid dynamics to express the speed of sound in
terms of thermodynamic quantities. Using the Gibbs-Duhem relation
$dP=sd\mathcal{T}$ one finds
\beq
\frac{dP}{d\vep}=s\frac{d\mathcal{T}}{d\vep}=\frac{s}{c_v}\,,
\eeq
where $c_v$ is the isovolumetric specific heat. So the black brane is
dynamically unstable to long wavelength, hydrodynamical perturbations,
if and only if it is locally
thermodynamically unstable, $c_v<0$. The `correlated stability
conjecture' of Gubser and Mitra \cite{Gubser:2000mm} posits precisely this
type of connection between dynamical
and thermodynamic stability.
The blackfold
method not only shows very simply why it holds for
hydrodynamic modes, but it also gives a
quantitative expression for the
unstable frequency in terms of local thermodynamics as
\beq\label{omegascv}
\Omega=\sqrt{\frac{s}{-c_v}}\;k+O(k^2)\,.
\eeq

\subsection{Viscous damping}

The previous analysis of the sound-wave instability employed the stress-energy tensor
of eq.~\eqref{blackTab}, which gives the perfect fluid approximation to the
intrinsic dynamics of the black brane. This tensor was obtained from the
stationary metric of the black brane. If we perturb this
brane, it will vibrate in its quasinormal modes, with damped
oscillations. The stress-energy tensor measured at large distance $r\gg
r_0$ from the black $p$-brane will reflect this damping through the appearance of
dissipative terms, proportional to derivatives of $u^a$ (the derivatives
of $r_0$ are proportional to these), so that
\begin{equation}\label{Tdiss}
T_{ab}=T^{(\mathit{perfect})}_{ab}-\zeta \theta P_{ab}-2\eta
\sigma_{ab} + O(D^2)\,.
\eeq
Here the orthogonal projector, expansion and shear of the velocity
congruence are
\beqa
P_{ab}&=&\eta_{ab}+u_a u_b\,,\qquad
\theta=D_a u^a\,,\\ 
\sigma_{ab}&=&{P_a}^c{P_b}^d\left(D_{\left(c\right.}u_{\left.d\right)}-
\frac{1}{p}\theta P_{cd}\right)\,,
\eeqa
and the coefficients $\eta$ and $\zeta$ are the effective shear and
bulk viscosity of the black brane.
They can be computed from a perturbative calculation very similar
to those in the context of the fluid/AdS-gravity
correspondence of \cite{chap:fluidgrav}. For the neutral black brane in
asymptotically flat space this calculation has been
carried out
in \cite{Camps:2010br}, with the result that
\beq\label{visco}
\eta=\frac{s}{4\pi}\,,\qquad \zeta=\frac{s}{2\pi}\left(\frac{1}{p}-v_s^2\right)\,,
\eeq
where $s$
is the entropy density of the black brane \eqref{locs} and $v_s$ the
speed of sound \eqref{soundspeed}.

Now we can solve the fluid equations \eqref{intreqs} for linearized
sound-mode
perturbations of the viscous fluid, and obtain the leading corrections
to the dispersion relation \eqref{Omk} at order $k^2$. 
For the black
brane fluid the result is
\beq\label{GLvisc}
\Omega=\frac{k}{\sqrt{n+1}}\left(1-\frac{n+2}{n\sqrt{n+1}}\,k r_0\right)\,,
\eeq
which is valid up to corrections $O(k^3)$. We see that viscosity has
the expected effect of damping the sound waves.
\begin{figure}
\includegraphics[scale=.9]{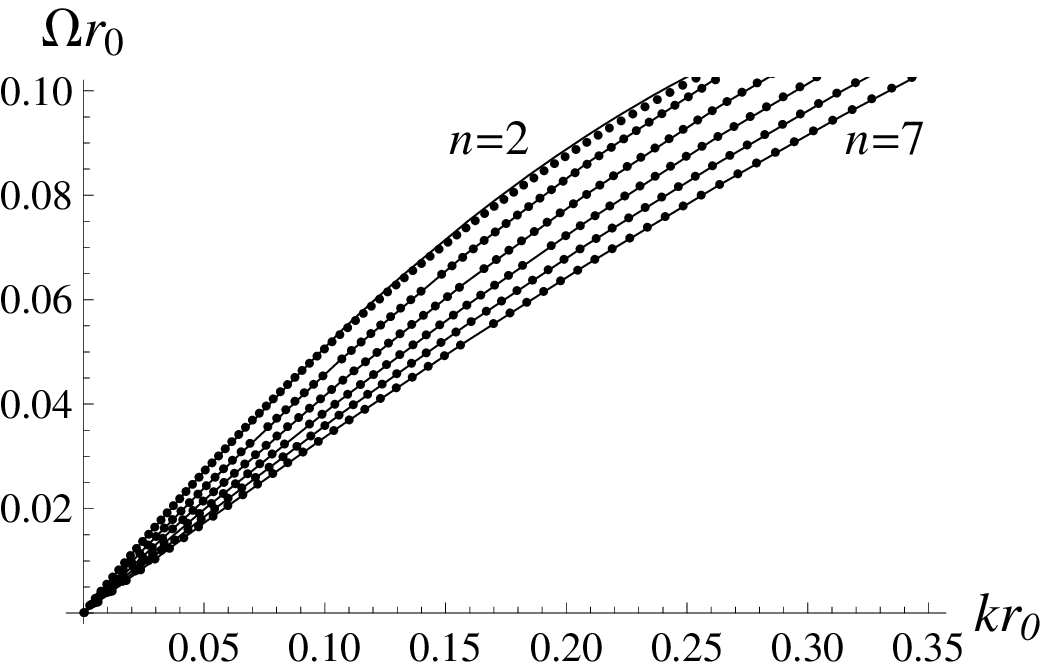}
\includegraphics[scale=.9]{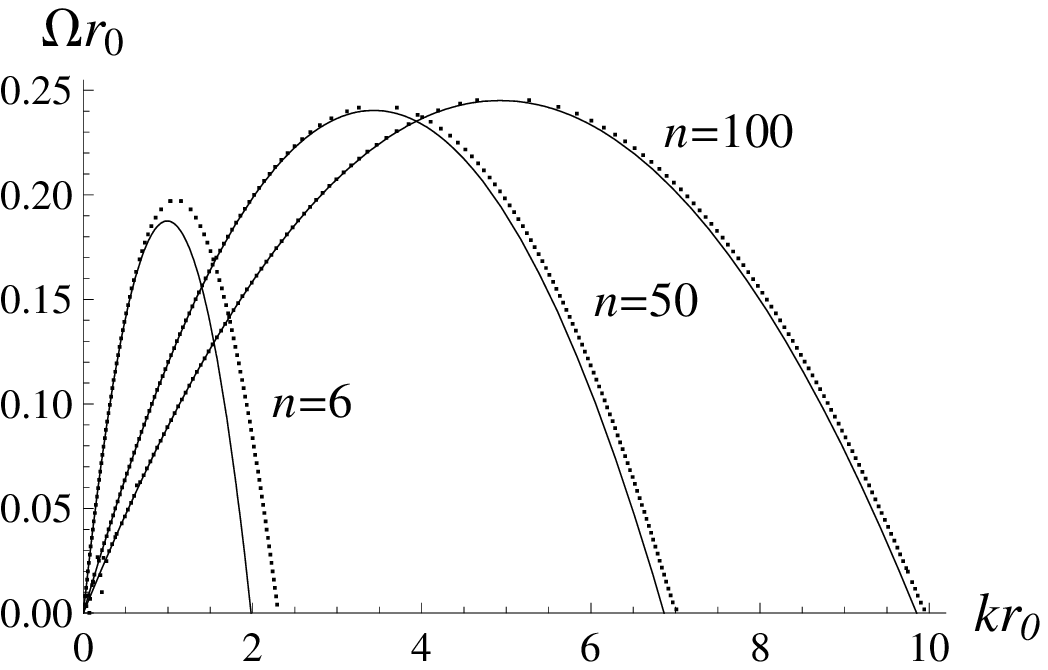}
\caption{$\Omega(k)$ for unstable modes of black
$p$-branes, in units of $1/r_0$ (the results depend only on
$n$ in \eqref{defn}).
The continuous curves are the analytical approximation
\eqref{GLvisc}, the dots are the numerical solution of the perturbations
of black branes. The top diagram shows results for $n=2,\dots,7$ at
small $k$. The bottom diagram shows the curves for $n=6,50,100$ at all $k$.
For large $n$, eq.~\eqref{GLvisc} underestimates the wavenumber of the
threshold mode by only $1/n$.
}\label{fig:compGL}
\end{figure}
Figure~\ref{fig:compGL} compares this
dispersion relation to the numerical results obtained by solving the
linearized perturbations of a black string.

Eq.~\eqref{GLvisc} gives excellent agreement to the numerical data for small $k
r_0$, but it also shows a remarkable overall resemblance to
them even when $kr_0$ is of order one, which is beyond the expected
range of validity of the approximation. The quantitative agreement
improves with increasing $n$: in an expansion in $1/n$,
eq.~\eqref{GLvisc} obtains the exact leading-order value for the
threshold wavenumber $k_{GL}\to \sqrt{n}/r_0$.
Ref.~\cite{Camps:2010br} suggests to explain this surprising agreement
by noting that the thermal wavelength $\lambda_T=1/T\sim r_0/n$ shrinks
to zero as $n\to\infty$ for fixed $r_0$. Quite plausibly, this effect
extends the range of wavelengths that fall under the remit of fluid
dynamics.

Let us emphasize how little has gone into the derivation of
\eqref{GLvisc}: only the equation of state $P(\vep)$ and the viscosities
$\eta$ and $\zeta$ --- actually, for a black string there is only
$\zeta$. The determination of these coefficients requires a study of
perturbations of the black string, but this can be carried out
analytically for all $n$ and $p$ and needs to be done only once.
Furthermore, the result for $\eta$ is known to be universal for black
holes, and the value of $\zeta$ saturates a proposed bound
\cite{Buchel:2007mf} which may plausibly be proven in generality. If there
exists such a general argument for the value of $\zeta$ for a black
brane, then the entire expression for the curve \eqref{GLvisc} can be
obtained, using simple algebra, from knowledge of only $dP/d\vep$.

Thus, the effective viscous fluid of the blackfold approach seems to
capture in a strikingly simple manner some of the most characteristic
features of black brane dynamics. This is a significant simplification
of the complexity of Einstein's equations.

\section{Extensions}

Although in section \ref{sec:bhsbfs} we have only considered Minkowski backgrounds,
the effective theory of blackfolds can be readily applied to the
construction of black holes in curved backgrounds, such as deSitter or
anti-deSitter spacetime with cosmological constant $\Lambda$. In this
case it is necessary that the thickness $r_0$ be much smaller than the
curvature radius $|\Lambda|^{-1/2}$, so the vacuum black brane
solution can be a good approximation in the near-zone. Black
rings, odd-spheres, and other blackfolds with characteristic radii $R$
that can be
larger or smaller than $|\Lambda|^{-1/2}$, are easy to construct in
these spacetimes \cite{Caldarelli:2008pz,Armas:2010hz}, as well as in other
non-trivial backgrounds such as Kaluza-Klein monopoles \cite{Camps:2008hb}.

Black $p$-branes can also carry on their worldvolumes the charges of
$q$-branes, $0\leq q\leq p$, which are sources of $(q+2)$-form gauge
field strengths $F_{q+2}$ \cite{Caldarelli:2010xz,bfss}. Then the worldvolume fluid includes
a conserved $q$-brane number current. For $q=0$ this is a theory of an
isotropic fluid with a conserved particle number, but when $q\geq 1$ the
brane current makes the fluid anisotropic. 

Blackfolds with a spatially compact worldvolume that supports these
currents correspond to black holes that source the field $F_{q+2}$. When
$q=0$ they carry a conserved charge of a Maxwell field. When $q\geq 1$
they carry a dipole of the field. In contrast to neutral blackfolds,
black holes constructed as charged blackfolds need not be ultraspinning:
the rotation may be small if the charge of the black brane is close to
its upper extremal limit. The presence of charge close to extremality
can also eliminate, in certain cases, the Gregory-Laflamme instability
of the black brane. Then, the black holes that result may be dynamically stable.

Of particular interest for string theory are blackfolds that carry
Ramond-Ramond charges. In general, blackfold techniques are an
appropriate tool for the study of configurations of D-branes in the
probe approximation, in the case that the D-brane worldvolume theory has
a thermal population of excitations. The blackfold gives a gravitational
description of this thermally excited worldvolume, with a horizon that
on short scales is like that of the straight black D-brane. Like in the
AdS/CFT correspondence, this gravitational description of the
worldvolume theory is appropriate when there is a stack of a large
number of D-branes (although not so large as to cause a strong
backreaction on the background) and the theory is strongly coupled.
Ref.~\cite{Grignani:2010xm} develops these methods to study a thermal
version of the D3-brane bion.

\section*{Acknowledgments}

I am indebted to my collaborators in the development of the blackfold
approach: Marco Caldarelli, Joan Camps, Nidal Haddad, Troels Harmark,
Vasilis Niarchos, Niels Obers, Mar{\'\i}a J.~Rodr{\'\i}guez. I also thank Pau
Figueras for the numerical data used in figure~\ref{fig:compGL} Work
supported by MEC FPA2010-20807-C02-02, AGAUR 2009-SGR-168 and CPAN
CSD2007-00042 Consolider-Ingenio 2010

\appendix
\renewcommand*\thesection{Appendix:}
\renewcommand*\thesubsection{\Alph{section}.\arabic{subsection}}
\renewcommand{\theequation}{A.\arabic{equation}}

\section{Geometry of submanifolds}
\setcounter{equation}{0}  
\subsection{Extrinsic curvature}

For a submanifold $\mathcal{W}$ embedded as $X^\mu(\sigma^a)$,
the pull-back of the spacetime metric onto $\mathcal{W}$, $\gamma_{ab}$, is
\eqref{gammaalbe} and the first fundamental form of the surface,
$h_{\mu\nu}$, is
obtained as
eq.~\eqref{defhmn}. It satisfies 
\beq\label{projdx}
{h^\mu}_\nu\partial_a X^\nu=\partial_a X^\mu\,,\qquad
h^\mu{}_\nu h^\nu{}_\rho =h^\mu{}_\rho\,,
\eeq
so $h^\mu{}_\nu$ projects tensors onto directions tangent to
$\mathcal{W}$. $\perp_{\mu\nu}$, introduced in \eqref{orthoproj},
projects onto orthogonal directions, 
\beq\label{projdx2}
\perp_{\mu\nu}\partial_a X^\mu=0\,,\qquad 
{\perp_\mu}^\nu {\perp_\nu}^\rho ={\perp_\mu}^\rho
\,.
\eeq
The shape of the embedding of the submanifold $\mathcal W$ is captured
by the second fundamental tensor, or extrinsic curvature tensor,
\eqref{Kext}.
Symmetry of the first two indices of ${K_{\mu\nu}}^\rho$ is
equivalent to the integrability of the subspaces orthogonal to
${\perp_\mu}^\nu$. To see this, let $s$ and $t$
be any two vectors in this subspace,
\beq
{\perp^\mu}_\nu t^\nu=0\,,\qquad {\perp^\mu}_\nu s^\nu=0\,. 
\eeq
Then one can easily prove from the definition of ${K_{\mu\nu}}^\rho$ that 
\beq
s^\mu t^\nu {K_{\mu\nu}}^\rho={\perp^\rho}_\mu \nabla_{s} t^\mu\,, 
\eeq
so
\beq
{K_{[\mu\nu]}}^\rho=0 \;\Leftrightarrow\; 
0={\perp^\rho}_\mu (\nabla_{s} t^\mu-\nabla_{t} s^\mu) = {\perp^\rho}_\mu [s,t]^\mu \,.
\eeq
The vanishing of the last commutator is equivalent, though Frobenius'
theorem, to the integrability of the subspace orthogonal to
${\perp_\mu}^\nu$. Therefore the extrinsic curvature tensor of the
submanifold $\mathcal W$ satisfies ${K_{[\mu\nu]}}^\rho=0$.

Now let $N$ be any vector orthogonal to $\mathcal{W}$. Then
\beq\label{NK}
N_\rho {K_{\mu\nu}}^\rho=N_\rho {h_\nu}^\sigma \oln_\mu {h_\sigma}^\rho=
-{h_\nu}^\rho \oln_\mu N_\rho\,.
\eeq

Background tensors ${t_{\mu_1 \mu_2\dots}}^{\nu_1\nu_2\dots}$ can be
pulled-back onto worldvolume tensors ${t_{a_1
a_2\dots}}^{b_1 b_2\dots}$ using
$\partial_a X^\mu$ as
\beq\label{wvtensors}
{t_{a_1 a_2\dots}}^{b_1 b_2\dots}=
\partial_{a_1} X^{\mu_1}\partial_{a_2} X^{\mu_2} \cdots
\partial^{b_1} X_{\nu_1}\partial^{b_2} X_{\nu_2}\cdots
{t_{\mu_1 \mu_2\dots}}^{\nu_1\nu_2\dots}\,,
\eeq
where
\beq
\partial^{b} X_{\nu}=\gamma^{bc}h_{\nu\rho}\partial_{c} X^{\rho}\,.
\eeq

Even when ${t_{\mu_1 \mu_2\dots}}^{\nu_1\nu_2\dots}$ has all indices parallel to $\mathcal{W}$, in general $\oln_\mu
{t_{\mu_1 \mu_2\dots}}^{\nu_1\nu_2\dots}$ has both parallel and
orthogonal components. The parallel projection along all indices is
related to the worldvolume covariant derivative $D_a{t_{a_1
a_2\dots}}^{b_1 b_2\dots}$ as in \eqref{wvtensors}. 
Then, the divergences of background and worldvolume tensors are related as
\beq\label{divs}
{h^{\nu_1}}_{\mu_1}\cdots \oln_\rho t^{\rho{\mu_1}\dots}
=\partial_{a_1}X^{\nu_1}\cdots D_c t^{c a_1\dots}\,.
\eeq

\subsection{Variational calculus}

Consider a congruence of curves
with tangent vector $N$, that
intersect
$\mathcal{W}$ orthogonally
\beq
N^\mu h_{\mu\nu}=0\,,\qquad N^\mu \perp_{\mu\nu}=N_\nu\,,
\eeq 
and Lie-drag $\mathcal{W}$ along these curves. 
The congruence is arbitrary, other than
requiring it to be smooth in a finite
neighbourhood of $\mathcal{W}$, so this realizes arbitrary smooth deformations
of the worldvolume $X^\mu \to X^\mu +N^\mu$.

Consider now the Lie derivative of $h_{\mu\nu}$ along $N$. In general,
\beq
\Ld_N h_{\mu\nu}=N^\rho \nabla_\rho h_{\mu\nu}+h_{\rho\nu}\nabla_\mu N^\rho
+h_{\mu\rho}\nabla_\nu N^\rho\,.
\eeq
Using \eqref{NK} one can derive
\beq
N_\rho {K_{\mu\nu}}^\rho=
-\frac{1}{2}{h_\mu}^\lambda {h_\nu}^\sigma \Ld_N h_{\lambda\sigma}\,.
\eeq
Taking the trace,
\beq
N_\rho K^\rho=-\frac{1}{2}h^{\mu\nu} \Ld_N h_{\mu\nu}=
-\frac{1}{\sqrt{|h|}}\Ld_N \sqrt{|h|}\,,
\eeq
where $h=\det h_{\mu\nu}$. These equations generalize well-known
expressions for the extrinsic curvature of a codimension-1 surface. 

Consider now a functional of the embedding of the form
\beq
I=\int_{\mathcal{W}} \sqrt{|h|}\, \Phi
\eeq
where $\Phi$ is a worldvolume function. Then
\beq
\delta_N I=\Ld_N\left(\sqrt{|h|}\,\Phi\right)=
\sqrt{|h|}\left(-N_\rho K^\rho \Phi+N^\rho\partial_\rho\Phi\right)\,.
\eeq
Since $N$ is an arbitrary orthogonal vector we have
\beq\label{varI}
\delta_N I=0 \quad 
\Leftrightarrow \quad K^\rho =\perp^{\rho\mu}\partial_\mu \ln \Phi\,.
\eeq
If $\Phi$ is constant then we recover the equation $K^\rho=0$ for minimal-volume submanifolds.

\endappendix

\end{document}